\documentclass[final,5p,times,twocolumn,authoryear]{elsarticle}

\usepackage{graphicx}
\usepackage{natbib}
\usepackage[colorlinks=true]{hyperref}
\usepackage{txfonts}
\usepackage{subfigure}
\usepackage{float}
\usepackage{bigints}
\usepackage{twoopt}
\usepackage{xspace}
\usepackage{balance}
\usepackage{upgreek}
\usepackage{multirow}
\usepackage{lineno}
\usepackage[procnames]{listings}
\usepackage{amssymb}
\usepackage{amsmath}
\usepackage[svgnames]{xcolor}

\journal{Astronomy and Computing}

\definecolor{keywords}{RGB}{255,0,90}
\definecolor{listinggray}{gray}{0.9}
\definecolor{lbcolor}{rgb}{0.9,0.9,0.9}

\lstdefinestyle{C++} {
  language=C++,
  backgroundcolor=\color{lbcolor},
  basicstyle=\scriptsize\upshape\ttfamily,
  commentstyle=\color{blue},
  classoffset=1,
  morekeywords={cosmobl, DirCosmo, DirLoc, Cosmology, par, Object, Galaxy, Catalogue, random_catalogue_box, TwoPointCorrelation},
  keywordstyle=\color{ForestGreen},
  classoffset=2,
  morekeywords={fsigma8, nObjects, setParameters, measure_xi, write_xi},
  keywordstyle=\color{red},
  classoffset=0,
  tabsize=2,
  captionpos=b,
  frame=lines,
  frameround=fttt,
  numbers=left,
  numberstyle=\tiny,
  numbersep=5pt,
  breaklines=true,
  showstringspaces=false
}

\lstdefinestyle{Python} {
  language=Python,
  backgroundcolor=\color{lbcolor},
  basicstyle=\scriptsize\upshape\ttfamily,
  commentstyle=\color{blue},
  classoffset=1,
  morekeywords={Cosmology, pyCosmologyCBL},
  keywordstyle=\color{ForestGreen},
  classoffset=2,
  morekeywords={D_C},
  keywordstyle=\color{red},
  classoffset=0,
  tabsize=2,
  captionpos=b,
  frame=lines,
  numbers=left,
  numberstyle=\tiny,
  numbersep=5pt,
  breaklines=true,
  showstringspaces=false,
  procnamekeys={def,class}
}

\newcommand{\DM}{\mathrm{DM}}

\begin{document}

\lstset{language=C++}

\begin{frontmatter}

  \title{CosmoBolognaLib: C++ libraries for cosmological
    calculations}
  
  \author[add1,add2,add3]{Federico Marulli}
  \ead{federico.marulli3@unibo.it}
  \ead[url]{https://www.unibo.it/sitoweb/federico.marulli3}
  \author[add1]{Alfonso Veropalumbo}
  \author[add1,add2]{Michele Moresco}
  \address[add1]{Dipartimento di Fisica e Astronomia - Universit\`{a} di Bologna, viale Berti Pichat 6/2, I-40127 Bologna, Italy}
  \address[add2]{INAF - Osservatorio Astronomico di Bologna, via Ranzani 1, I-40127 Bologna, Italy}
  \address[add3]{INFN - Sezione di Bologna, viale Berti Pichat 6/2, I-40127 Bologna, Italy}


  \begin{abstract}
    We present the {\small CosmoBolognaLib}, a large set of Open
    Source C++ numerical libraries for cosmological calculations.
    {\small CosmoBolognaLib} is a {\em living project} aimed at
    defining a common numerical environment for cosmological
    investigations of the large-scale structure of the Universe. In
    particular, one of the primary focuses of this software is to help
    in handling astronomical catalogues, both real and simulated,
    measuring one-point, two-point and three-point statistics in
    configuration space, and performing cosmological analyses. In this
    paper, we discuss the main features of this software, providing an
    overview of all the available C++ classes implemented up to
    now. Both the {\small CosmoBolognaLib} and their associated
    {\small doxygen} documentation can be freely downloaded at \url
    {https://github.com/federicomarulli/CosmoBolognaLib} . We
    provide also some examples to explain how these libraries can be
    included in either C++ or Python codes.
  \end{abstract}
  

\end{frontmatter}


\section{Introduction}
\label{sec:intro}

Numerical tools for cosmological calculations are one of the crucial
ingredients in the increasingly ambitious investigations of the
large-scale structure of the Universe.  Several public libraries for
astronomical calculations are nowadays available, in different
languages, such as e.g. {\small CfunBASE} \citep{taghizadeh-popp2010},
{\small CosmoPMC} \citep{kilbinger2011}, {\small AstroML}
\citep{astroML2012}, {\small CUTE} \citep{alonso2012}, {\small
  Astropy} \citep{astropy2013}, {\small Cosmo++} \citep{aslanyan2014},
      {\small CosmoloPy}\footnote{http://roban.github.com/CosmoloPy/},
      {\small NumCosmo} \citep{numcosmo2014}, {\small TreeCorr}
      \citep{jarvis2015}.

Aiming at defining a common environment for handling extragalactic
source catalogues, performing statistical analyses and extracting
cosmological constraints, we implemented a large set of C++ libraries,
called {\small CosmoBolognaLib} (hereafter CBL), specifically focused
on numerical computations for cosmology, thus complementing the
available software. In particular, the CBL provide highly optimised
algorithms to measure two-point (2PCF) and three-point correlation
functions (3PCF), exploiting a specifically designed parallel
chain-mesh algorithm to count pairs and triplets. Several types of
correlation functions can be computed, such as the angle-averaged
2PCF, the 2D 2PCF in both Cartesian and polar coordinates and its
multipole moments, the angular, projected and deprojected 2PCF, the
clustering wedges, the filtered 2PCF, and the connected and reduced
3PCF (see \S \ref{subsec:measurements}).  Moreover, a large set of
methods are provided to construct random catalogues, to estimate
errors and to extract cosmological constraints from clustering
analyses (see \S \ref{subsec:models}). These features represent the
main novelty of the presented libraries.

The CBL are fully written in C++. They can be included either in C++
codes or, alternatively, in high-level scripting languages through
wrapping. We provide an example code that shows how to include the
CBL in Python scripts in \ref{subsec:ex_python}.

This effort can be considered as a {\em living project}, started a few
years ago and intended to be continued in the forthcoming years. The
following is the list of scientific publications that have been fully
or partially performed using the presented libraries:
\citet{marulli2011, marulli2012a, marulli2012b, marulli2013,
  marulli2015, giocoli2013, villaescusa2014, moresco2014,
  veropalumbo2014, veropalumbo2015, sereno2015, moresco2015,
  petracca2015}. Thanks mainly to the adopted object-oriented
programming technique, the CBL are flexible enough to be significantly
extented.

In this paper, we present the main features of the current version of
the CBL, that is fully publicly available\footnote{
  \url{https://github.com/federicomarulli/CosmoBolognaLib} and \\ \url
      {http://apps.difa.unibo.it/files/people/federico.marulli3}},
together with the documentation obtained with {\small
  doxygen}\footnote{ \url{www.doxygen.org}}. A set of sample codes,
that explain how to use these libraries in either C++ or Python
software, is provided at the same webpage.

The paper is organised as follows. In \S\ref{sec:cosmo} we describe
the CBL class for cosmological computations. In \S\ref{sec:cat} we
present the classes implemented for handling catalogues of
extragalactic sources. 2PCF and 3PCF can be measured and modelled with
specific classes that are described in
\S\ref{sec:clustering}. \S\ref{sec:stats} presents the CBL methods for
statistical analyses. In \S\ref{sec:func} we provide a brief
description of the other CBL functions used for several generic
calculations. Finally, in \S\ref{sec:conc} we draw our
conclusions. Compiling instructions and a few sample codes are
reported in \ref{sec:compile} and \ref{sec:examples}, respectively.


\section{Cosmology}
\label{sec:cosmo}

All the cosmological functions defined in the CBL are implemented as
public members of the class {\mbox{\color{ForestGreen}
    cosmobl::Cosmology}}\footnote{cosmobl is the global namespace of
  the CBL.}. The private parameters of this class are the following:
the matter density, that is the sum of the density of baryons, cold
dark matter and massive neutrinos (in units of the critical density)
at z=0, $\Omega_{\rm matter}$; the density of baryons at z=0,
$\Omega_{\rm baryon}$; the density of massive neutrinos at z=0,
$\Omega_{\nu}$; the effective number of relativistic degrees of
freedom, $N_{\rm eff}$; the number of massive neutrino species; the
density of dark energy at z=0, $\Omega_{\rm DE}$; the density of
radiation at z=0, $\Omega_{\rm radiation}$; the Hubble parameter,
$h=H_0/100$; the initial scalar amplitude of the power spectrum,
$A_s$; the primordial spectral index, $n_{\rm spec}$; the two
parameters of the dark energy equation of state in the
Chevallier-Polarski-Linder parameterisation \citep{chevallier2001,
  linder2003}, $w_0$ and $w_a$; the non-Gaussian amplitude, $f_{\rm
  NL}$; the non-Gaussian shape -- local, equilateral, enfolded,
orthogonal \citep{fedeli2011}; the model used to compute distances
(used only for some specific interacting dark energy models, see
\citealt{marulli2012a}); a variable called {\em unit}, used to choose
between physical units or cosmological units (that is in unit of
$h$). If the above parameters are not specified when creating an
object of this class, default values from Planck cosmology will be used
\citep{planck2014}. In any case, each cosmological parameter can be
set individually, when required.

Once the cosmological model has been chosen by setting the parameters
described above, a large set of cosmological functions can then be
used. We provide here a brief overview of the main functions of the
class. The full explanation of the whole set of class members can be
found in the {\small doxygen} documentation at the CBL webpage.

Several functions are available to estimate the redshift evolution of
all the relevant cosmological parameters, to compute the lookback and
cosmic times, to estimate cosmological distances and volumes, and to
convert redshifts into comoving distances and viceversa.

There are methods to estimate the number density and mass function of
dark matter haloes. Specifically, the code implements the
  following equation \citep[see e.g.][]{marulli2011}:
  \begin{equation}
    \frac{M dM}{\bar \rho}\frac{dn(M,z)}{dM} = \zeta f(\zeta)
    \frac{d\zeta}{\zeta} \, ,
    \label{eq:MF}
  \end{equation}
  with $\zeta \equiv [\delta_{\rm sc}(z)/\sigma(M)]^2$, where
  $\delta_{\rm sc}(z)$ is the overdensity required for spherical
  collapse at $z$, $\bar \rho = \Omega_{\rm matter} \rho_c$, $\rho_c$
  is the critical density of the Universe, and $dn(M,z)$ is the halo
  number density in the mass interval $M$ to $M + dM$.  The variance
  of the linear density field is given by
  \begin{equation}
    \sigma^2(M) = \int dk \frac{k^2 P_{\rm lin}(k)}{2 \pi^2} |W(kR)|^2 \,
    ,
  \end{equation}
  where the top-hat window function is $W(x) = (3/x^3)(\sin x - x \cos
  x)$, with $R = (3M/4\pi \bar \rho)^{1/3}$.  At the moment, the
  implemented mass function models are the following:
  \citet{press1974, sheth1999, jenkins2001, warren2006, shen2006,
    reed2007, pan2007, tinker2008, angulo2012}.

Methods to estimate the effective linear bias of dark matter haloes
are provided as well. The effective bias is computed through the
  following integral:
  \begin{equation}
    b(z) = \frac{\int_{M_{\rm min}}^{M_{\rm max}} n(M,z) b(M,z)dM}{\int_{M_{\rm min}}^{M_{\rm max}} n(M,z)dM} \, ,
    \label{eq:bias} 
  \end{equation}
  where b(M,z) is the linear bias and $n(M,z)$ is the halo number
  density.  The available parameterisations are: \citet{sheth1999,
    sheth_mo_tormen2001, tinker2010}.

A large set of functions is provided to estimate the real-space and
redshift-space power spectra and 2PCF (see
\S\ref{subsec:ex_cosmology}), and to assess the cosmic mass accretion
history \citep{giocoli2013}.  To estimate the dark matter power
spectrum and all the derived quantities, such as the mass variance
used to compute the mass function and bias, the user can choose
between one of the following external codes: {\small CAMB}
\citep{lewis2000}, {\small MPTbreeze} \citep{crocce2012}, {\small
  CLASS} \citep{lesgourgues2011, blas2011}, Eisenstein\&Hu code
\citep{eisenstein1998, eisenstein1999}. The latter will be exploited
authomatically by the CBL via specific functions used to set the
parameter files conveniently.


\section{Catalogues}
\label{sec:cat}

The class {\mbox{\color{ForestGreen} cosmobl::Catalogue}} is used to
handle samples of astronomical objects. The present version of the CBL
provides specific classes for galaxies, clusters of galaxies, dark
matter haloes and generic mock objects. However, the code structure is
sufficiently versatile to easily include new objects or to extend the
present ones, e.g. by adding new properties. Once the catalogue is
created, several operations can be performed, such as estimating the
distribution of any property of the object members, dividing the
catalogues in sub-samples, or creating a smoothed version of the
original catalogue. Moreover, a catalogue can be passed to other
objects as an input, e.g. to estimate 2PCF and 3PCF (see
\S\ref{subsec:measurements}), or to assess errors through the
jackknife or boostrap techniques (see
\S\ref{subsec:errors}). Catalogues can also be added together, or they
can be enlarged by adding new single objects. 

For a fast spatial search of objects in the catalogues, we implemented
a highly optimised chain-mesh method, specifically designed for
counting object pairs and triplets in a specified range of scales. The
algorithm implements a pixelization scheme, similar to the one
described in \citet{alonso2012}. First, the catalogue is divided into
cubic cells, and the indexes of all the objects in each cell are
stored in vectors. Then, to find all the objects close to a given one,
the search is performed only on the cells in the chosen scale range,
thus minimizing the amount of useless counts of objects at too large
separations. In this way, the efficiency of the method depends
primarily on the ratio between the scale range of the searching region
and the maximum separation between the objects in the catalogue. This
is particularly useful when measuring 2PCF and 3PCF (see \S
\ref{sec:clustering}). For alternative searching algorithms, such as
kd-tree and ball-tree methods, see e.g. \citet{jarvis2015}.

The chain-mesh method is implemented in the four classes:
{\mbox{\color{ForestGreen} cosmobl::ChainMesh}},
{\mbox{\color{ForestGreen} cosmobl::ChainMesh1D}},
{\mbox{\color{ForestGreen} cosmobl::ChainMesh2D}} and
{\mbox{\color{ForestGreen} cosmobl::ChainMesh3D}}, designed to handle
chains in 1, 2 and 3 dimensions. An example that shows how to create
and use objects of these classes is provided in \ref{subsec:ex_2pt}
and at the CBL webpage.


\section{Clustering}
\label{sec:clustering}

One of the main focuses of the CBL is to provide functions to measure
and model the clustering properties of astronomical sources. In this
section, we present a general description of the main features of the
current version of CBL methods for clustering analyses.


\subsection{Measurements}
\label{subsec:measurements}

Two CBL classes, {\mbox{\color{ForestGreen}
    cosmobl::TwoPointCorrelation}} and {\mbox{\color{ForestGreen}
    cosmobl::ThreePointCorrelation}}, can be used to measure 2PCF and
3PCF, respectively. The 2PCF, $\xi(r)$, is implicitly defined as
$dP_{12} = n^2[1+\xi(r)]dV_1dV_2$, where $n$ is the average number
density, and $dP_{12}$ is the probability of finding a pair with one
object in the volume $dV_1$ and the other one in the volume $dV_2$,
separated by a comoving distance $r$. To estimate this function, the
CBL provide an implementation of the \citet{landy1993} estimator:
\begin{equation}
  \xi(r) = \frac{DD(r)+RR(r)-2DR(r)}{RR(r)} \, ,
  \label{eq:xiLS}
\end{equation}
where $DD$, $RR$ and $DR$ are the data-data, random-random and
data-random normalised pair counts, respectively, for a separation
bin $r\pm\mathrm{d}r/2$. The following types of clustering functions
can be computed:
\begin{itemize}
\item the angle-averaged 2PCF, $\xi(r)$;
\item the 2D 2PCF in both Cartesian and polar coordinates,
  $\xi(r_p,\pi)$ and $\xi(r,\mu)$, that is as a function of
  perpendicular, $r_p$, and parallel, $\pi$, line-of-sight
  separations, and as a function of absolute separation,
  $r=\sqrt{r_p^2+\pi^2}$, and the cosine of the angle between the
  separation vector and the line of sight,
  $\mu\equiv\cos\theta=r_\parallel/r$, respectively;
\item the projected 2PCF:
  \begin{equation}
    w(r_p)=2\int^{\pi_{\rm max}}_{0} d\pi' \,
    \xi(r_p, \pi') \, ;
    \label{eq:proj2pcf}
  \end{equation}
\item the deprojected 2PCF:
  \begin{equation}
    \xi(r) = -\frac{1}{\pi}\int^{r_{\rm max}}_r dr_p'
    \frac{dw_p(r_p')/dr_p}{\sqrt{r_p'^2-r^2}} \, ;
  \end{equation}
\item the angular 2PCF, $w(\theta)$, where $\theta$ is
  the angular separation;
\item the multipole moments of the 2PCF:
  \begin{equation}
    \xi_l(r) = \frac{2l+1}{2} \int_{-1}^{1} \xi(s,\mu) L_l(\mu) d\mu \, ,
    \label{eq:multipoles}
  \end{equation}
  where $L_l(\mu)$ is the Legendre polynomial of order $l$;
\item the clustering wedges \citep{kazin2012}:
  \begin{equation}
    \xi(r,\Delta \mu) \equiv \frac{\int_{\mu_{\rm min}}^{\mu_{\rm max}} \mathrm{d}\mu^{\prime} \xi(r,\mu^{\prime})} 
       {\int_{\mu_{\rm min}}^{\mu_{\rm max}} \mathrm{d}\mu^{\prime}} \, ;
       \label{eq:xiwedges}
  \end{equation}
\item the filtered correlation function \citep{xu2010, cervantes2012}:
  \begin{equation}
    \omega_0(r_s) = 4\pi\int_0^{r_s} \frac{\mathrm{d}r}{r_s} \left( \frac{r}{r_s} \right)^2 \xi(r)  
    W(r,r_s) \, ,
    \label{eq:filtxi}
  \end{equation}
  where the filter is:
  \begin{equation}
    \left\{
    \begin{array}{ll}
      W(x)= 4x^{2} (1-x)\left (\frac{1}{2}-x \right) & 0<x<1 \, ,   \\
      W(x)= 0      & \textrm{otherwise} \, ;
    \end{array}
    \right.
    \label{eq:xifilt}
  \end{equation}
  and $x\equiv (r/r_s)^3$.
\end{itemize}

The CBL provide also methods both to construct random catalogues with
different geometries and to read them from files, in case they have
been already computed. Specifically, there are functions for both
cubic and conic geometries, in order to construct random catalogues
both for cubic simulation snapshots, and for mock or real catalogues
in light-cones.

Analogously to the 2PCF, the 3PCF, $\zeta(r_{12},r_{23},r_{31})$, is
defined as
$dP_{123}=n^3[1+\xi(r_{12})+\xi(r_{23})+\xi(r_{31})+\zeta(r_{12},r_{23},r_{31})]
dV_{1}dV_{2}dV_{3}$, where $n$ is the average density of objects, and
$V_i$ are comoving volumes. It is calculated using the
\citet{szapudi1998} estimator:
\begin{equation}
\zeta(r_{12},r_{23},r_{31}) = \frac{DDD-3DDR+3DRR-RRR}{RRR} \, ,
\end{equation}
where $DDD$, $RRR$, $DDR$, and $DRR$ are the normalised numbers of
data triplets, random triplets, data-data-random triplets, and
data-random-random triplets, respectively. The algorithm fixes two
sides of the triangles and varies the angle, $\theta$, between them.
The 3PCF can be measured both in Cartesian coordinates,
$\zeta(r_{12},r_{23},r_{31})$, and as a function of the angle between
the triangle sides, $\zeta(\theta)$. Also the reduced 3PCF,
$Q(r_{12},r_{13},\theta)$, can be computed. The latter is defined as
follows:
\begin{equation}
Q = \frac{\zeta(r_{12},r_{23},\theta)}
{\xi(r_{12})\xi(r_{23})+\xi(r_{23})\xi(r_{31})+\xi(r_{31})\xi(r_{13})}
\, .
\end{equation}
The minimum and maximum separations used to count pairs and triplets
and the binning size are free parameters that can be set by the user.

The algorithms to measure the above clustering functions use the
chain-mesh method described in \S \ref{sec:cat}. The code exploits
also multithreaded parallelism. Specifically, all the loops to count
the number of object pairs and triplets are parallelized via
OpenMP\footnote{\url{http://openmp.org/wp/}}. The code performances
scale almost linearly with the number of threads.

All the operations related to pair and triplet counts are implemented
in the classes {\mbox{\color{ForestGreen} cosmobl::Pairs}},
{\mbox{\color{ForestGreen} cosmobl::Pairs2D}},
{\mbox{\color{ForestGreen} cosmobl::Pairs3D}},
{\mbox{\color{ForestGreen} cosmobl::Triplets}},
{\mbox{\color{ForestGreen} cosmobl::Triplets2D}} and
{\mbox{\color{ForestGreen} cosmobl::Triplets3D}}. These functions have
been deeply tested with both simulated catalogues \citep{marulli2011,
  marulli2012a, marulli2012b, marulli2015, villaescusa2014,
  moresco2014, petracca2015}, and real catalogues of galaxies
\citep{marulli2013, moresco2015} and galaxy clusters
\citep{veropalumbo2014, veropalumbo2015, sereno2015}.
\ref{subsec:ex_2pt} provides an example that shows how to create a
random catalogue and measure the 2PCF. Further examples can be found
at the CBL webpage.


\subsection{Errors}
\label{subsec:errors}
To estimate the errors in 2PCF measurements, the CBL provide specific
functions to estimate the covariance matrix defined as follows:
\begin{equation}
  C_{i,j} = \mathcal{F} \sum_{k=1}^{N}(\xi^k_i-\hat{\xi_i})
  (\xi^k_j-\hat{\xi_j}) \, ,
  \label{eq:jkcov}
\end{equation}
where the indexes $i$ and $j$ run over the spatial bins of the 2PCF,
the index $k$ refers to the 2PCF of the $k^{th}$ realisation,
$\hat{\xi}$ is the mean 2PCF over all the $N$ realisations. The factor
$\mathcal{F}$ is equal to either $(N-1)/N$ or $1/N$, in case of
jackknife or boostrap errors, respectively.  By definition, the
diagonal elements of this matrix are the variance of the i$-th$
spatial bin: $\sigma_i^2$.  The covariance matrix can be estimated
with three alternative methods \citep[see e.g.][]{norberg2009}:

\begin{itemize}

\item {\em analytic errors}: 2PCF errors can be estimated
  analytically, assuming Poisson statistics. The CBL contain functions
  to compute analytic errors used to set the diagonal elements of
  $C_{i,j}$;
  
\item {\em internal errors}: the CBL provide functions to estimate
  errors by sub-sampling the data catalogue and measuring the 2PCF for
  all but one region -- {\em jackknife}, or for a random extraction of
  regions -- {\em bootstrap}. The volume can be partitioned either in
  cubic sub-regions, useful e.g. when analysing simulation snapshots,
  or in sub-regions of generic geometry using the external software
  {\small MANGLE} to reconstruct the angular mask \citep{swanson2008};
  
\item {\em external errors}: the CBL class {\mbox{\color{ForestGreen}
    cosmobl::LogNormal}} can be used to generate lognormal mock
  catalogues \citep{coles1991}, with a specified power spectrum, from
  which the covariance matrix can be estimated.

\end{itemize}

Analogous methods for the 3PCF will be included in a forthcoming
version of the CBL.


\subsection{Models}
\label{subsec:models}

The modelling of the 2PCF is managed by the class
{\mbox{\color{ForestGreen} cosmobl::ModelTwoPointCorrelation}}, which
provides methods to model all the two-point statistics described in
\S\ref{subsec:measurements}. The following sections present an
overview of the available facilities.


\subsubsection{The angle-averaged 2PCF}

The angle-averaged 2PCF of cosmic tracers, $\xi(r)$, can be modelled
both in real space and in redshift space.  In real space, we implement
the simple model:
\begin{equation}
\xi(r) = b^2\xi_{\DM}(\alpha r) \, ,
\end{equation}
where $\xi(r)$ is the model 2PCF of the tracers,
$\xi_{\DM}(r)$ is the dark matter 2PCF (see \S\ref{sec:cosmo}), $b$ is
the linear bias of dark matter haloes or galaxies, and $\alpha$ is the
shift parameter defined as:
\begin{equation}
\alpha \equiv \frac{D_V}{r_s} \frac{r_s^{fid}}{D_V^{fid}} \, ,
\end{equation}
where $D_V$ is the isotropic volume distance and $r_s$ is the position
of the sound horizon at decoupling.  This quantity is used when
exploiting the {\em standard ruler technique} in case of baryon
acoustic oscillations (BAO) fitting \citep{veropalumbo2014}.

To model the angle-averaged 2PCF in redshift space, we compute the
Fourier anti-transform of the redshift-space power spectrum, that can
be written as follows:
\begin{equation}
P(k,\mu) \, = \, P_{\DM}(k)\left( b^2+f\mu^2 \right)^2
\exp\left(-k^2\mu^2\sigma^2 \right) \, ,
\label{eq:pkrsd}
\end{equation}
where $P_{\DM}(k)$ is the dark matter power spectrum (see
\S\ref{sec:cosmo}), $f$ is the linear growth rate of cosmic
structures, and $\sigma$ is a damping scale term introduced to
describe the effect of Gaussian redshift errors. Specifically, the
relation between $\sigma$ and the redshift error $\sigma_z$ is:
\begin{equation}
\sigma \, = \, \frac{c\sigma_z}{H(z)} \, ,
\end{equation}
where $c$ is the speed of light and $H(z)$ is the Hubble function at
redshift $z$.  Ignoring the non-linear damping term, the
angle-averaged 2PCF model reads:
\begin{equation}
\xi(s) \, = \,
\left[b^2+\frac{2}{3}\frac{f}{b}+\frac{1}{5}\left(\frac{f}{b}\right)^2\right]
\cdot \xi_{\DM}(\alpha s) \, ,
\end{equation}
while in the most general case it becomes:
\begin{equation}
\xi(s) \, = \, b^2\xi'(s) + b\xi''(s) +\xi'''(s) \, , 
\end{equation}
where $\xi'(s)$, $\xi''(s)$ and $\xi'''(s)$ are, respectively, the Fourier
anti-transforms of:
\begin{flalign}
P'(k)  &= P_{\DM}(k) \frac{\sqrt{\pi}}{2 k \sigma} \mathrm{erf}(k\sigma) \, ;&\\
P''(k) &= \frac{f}{(k\sigma)^3} P_\mathrm{DM}(k) \left[
\frac{\sqrt{\pi}}{2}\mathrm{erf}(k\sigma) -
k\sigma\exp(-k^2\sigma^2)\right]\, ;&\\
P'''(k) &=
\frac{f^2}{(k\sigma)^5}P_\mathrm{DM}(k) \left\{ \frac{3\sqrt{\pi}}{8}\mathrm{erf}(k\sigma) \right. & \nonumber \\ 
&\left. - \frac{k\sigma}{4}\left[2(k\sigma)^2+3\right]\exp(-k^2\sigma^2) \right\} \, .& 
\end{flalign}
$P'(k)$, $P''(k)$ and $P'''(k)$ are the terms obtained by integrating
Eq.~(\ref{eq:pkrsd}) along $\mu$.

The {\mbox{\color{ForestGreen} cosmobl::ModelTwoPointCorrelation}}
class contains also a model for the de-wiggled power spectrum, used to
describe non-linear damping effects on BAO. In this case, the
dependence on non-linear effects is explicit, via the $\Sigma_{NL}$
parameter \citep{eisenstein2007}.  The de-wiggled power spectrum for
the dark matter is:
\begin{equation}
  P_{\DM}(k) \, = \, \left[ P_{\rm lin}(k) - P_{nw}(k) \right]
  \mathrm{e}^{-k^2\Sigma_{NL}^2/2} + P_{nw}(k) \, ,
  \label{eq:pkmodel}
\end{equation}
where $P_{\rm lin}$ is the linear dark matter power spectrum, $P_{nw}$
is the linear dark matter power spectrum without BAO, and
$\Sigma_{NL}$ describes the non-linear damping effect. Then the 2PCF is
computed as the Fourier anti-transform of the power spectrum.

Finally, the CBL provide a simple empirical model used to recover the
BAO peak position. It approximates the 2PCF as a combination of a
power law and a Gaussian function to reproduce the BAO peak shape
\citep{smith2008}:
\begin{equation}
\xi(s) \, = \, \left(\frac{s}{s_0}\right)^{-\gamma} +
\frac{A}{\sqrt{2\pi\sigma^2}}\exp\left(-\frac{(s-s_m)^2}{2\sigma^2}\right) \, ,
\end{equation}
where $s_0$ and $\gamma$ are the power-law normalisation and slope,
while $A$, $\sigma$ and $r_m$ are the Gaussian normalisation factor, the
standard deviation and the mean, respectively.


\subsubsection{Projected correlation function}

The model for the projected correlation function relies on the
assumption that the integration of the 2PCF along the line of sight
cancels out redshift-space distorsion effects. The model is derived by
changing the variable of the integration in Eq.~(\ref{eq:proj2pcf})
from $\pi$ to $r \equiv \sqrt{\pi^2+r_p^2}$:
\begin{equation}
    w(r_p)=b^2\int^{r_{\rm max}}_{0} dr' \,
    \frac{\xi_{\DM}(r)}{\sqrt{r^2-r_p^2}} \, .
\end{equation}


\subsubsection{2D 2PCF}
\label{subsec:2D2PCF}
The CBL provide methods to model the redshift-space 2D 2PCF and its
multipole moments. The so-called {\em dispersion model} is currently
implemented. In the linear regime, the redshift-space 2PCF can be
written as follows:
\begin{equation} 
\xi^{\rm lin}(s,\mu) =
\xi_0(s)P_0(\mu)+\xi_2(s)P_2(\mu)+\xi_4(s)P_4(\mu) \; ,
\label{eq:ximodellin}
\end{equation}
where $P_l$ are the Legendre polynomials \citep{kaiser1987}. The
multipoles moments are:
\begin{subequations}
  \begin{align}
    \xi_0(s) & = \left(1+ \frac{2}{3}\beta + \frac{1}{5}\beta^2 \right)
    \cdot \xi(r)
    \label{eq:xi0_1} \\ 
    & =  \left[ (b\sigma_8)^2 + \frac{2}{3} f\sigma_8 \cdot b\sigma_8 +
      \frac{1}{5}(f\sigma_8)^2 \right] \cdot \frac{\xi_{\rm DM}(r)}{\sigma_8^2} \; ,
    \label{eq:xi0_2} 
  \end{align}
\end{subequations}

\begin{subequations} 
  \begin{align}
    \xi_2(s) &  = \left(\frac{4}{3}\beta +
    \frac{4}{7}\beta^2\right)\left[\xi(r)-\overline{\xi}(r)\right] 
    \label{eq:xi2_1} \\ 
    & = \left[\frac{4}{3}f\sigma_8 \cdot b\sigma_8 +
      \frac{4}{7}(f\sigma_8)^2\right]\left[\frac{\xi_{\rm
          DM}(r)}{\sigma_8^2}-\frac{\overline{\xi}_{\rm
          DM}(r)}{\sigma_8^2}\right] \; ,
    \label{eq:xi2_2} 
  \end{align}
\end{subequations}

\begin{subequations} 
  \begin{align}
    \xi_4(s) & = \frac{8}{35}\beta^2\left[\xi(r) +
      \frac{5}{2}\overline{\xi}(r)
      -\frac{7}{2}\overline{\overline{\xi}}(r)\right] 
    \label{eq:xi4_1} \\ 
    & = \frac{8}{35}(f\sigma_8)^2\left[ \frac{\xi_{\rm
          DM}(r)}{\sigma_8^2} + \frac{5}{2}\frac{\overline{\xi}_{\rm
          DM}(r)}{\sigma_8^2}
      -\frac{7}{2}\frac{\overline{\overline{\xi}}_{\rm DM}(r)}{\sigma_8^2}
      \right] \; ,
    \label{eq:xi4_2} 
  \end{align}
\end{subequations}
where the {\em barred} functions are:
\begin{equation} 
\overline{\xi}_{\rm DM}(r) \equiv \frac{3}{r^3}\int^r_0dr'\xi_{\rm
  DM}(r')r'{^2} \; ,
\label{eq:xi_}
\end{equation}
\begin{equation}
\overline{\overline{\xi}}_{\rm DM}(r) \equiv \frac{5}{r^5}\int^r_0dr'\xi_{\rm DM}(r')r'{^4} \; .
\label{eq:xi__} 
\end{equation}

To account for non-linear dynamics, the linearly-distorted correlation
function is then convolved with the distribution function of pairwise
velocities, $f(v)$, \citep{peacock1996, peebles1980}:
\begin{equation} 
 \xi(s_\perp, s_\parallel) = \int^{\infty}_{-\infty}dv
 f(v)\xi\left(s_\perp, s_\parallel - \frac{v(1+z)}{H(z)}\right)_{\rm
   lin} \; ,
\label{eq:ximodel}
\end{equation}
where the pairwise velocity $v$ is expressed in physical
coordinates. Both exponential and Gaussian functions can be used to
model the distribution function $f(v)$ \citep[see
  e.g.][]{marulli2012b}.

More accurate models for redshift-space distortions will be included
in a forthcoming version of the CBL.


\subsubsection{Other statistics}

The CBL provide methods to model both the wedges of 2D 2PCF and the
filtered 2PCF described in \S\ref{sec:clustering}. This can be done by
assuming a model for the redshift-space distorted $\xi(s,\mu)$ (see
\S\ref{subsec:2D2PCF}), and then using
Eqs.(\ref{eq:xiwedges})-(\ref{eq:filtxi}). Methods to model the
angular 2PCF and the 3PCF are not yet available and will be added in a
future version of the CBL.


\section{Tools for statistical analyses}
\label{sec:stats}

We implemented generic classes to model measured quantities and derive
cosmological constraints.  Input data and models are fully
customizable. The implementation is done in the classes:
{\mbox{\color{ForestGreen} cosmobl::Data}}, {\mbox{\color{ForestGreen}
    cosmobl::Model}} and {\mbox{\color{ForestGreen}
    cosmobl::Parameter}}. The class {\mbox{\color{ForestGreen}
    cosmobl::Chi2}} implements standard minimum $\chi^2$ fitting
techniques.  The CBL provide also Bayesian inference methods based on
the Bayes' theorem:
\begin{equation}
p(\vec{\theta}\, | \, \vec{X}) \, = \, \frac{p(\vec{X} \, | \, \vec{\theta}) \, p(\vec{\theta})} {p(\vec{X})} \, ,
\label{eq:bayes}
\end{equation}
where $\vec{X}$ are the data, $\vec{\theta}$ are the model parameters,
$p(\vec{\theta})$ is the prior probability distribution of the
parameters, and $p(\vec{X} \, | \, \vec{\theta})$ and $p(
\vec{\theta}\, | \, \vec{X} )$ are the likelihood function and the
parameter posterior probability distribution, respectively.  The CBL
provide methods to perform the Markov Chain Monte Carlo (MCMC)
likelihood sampling technique.  The latter consists in sampling a
target distribution using a correlated random walk: every step is
extracted after a trial that depends only on the previous one (Markov
process). The steps are collected in chains, that define marginalised
posterior probability of the model parameters $p(\vec{\theta}\, | \,
\vec{X})$. We implemented two MCMC algorithms:
\begin{itemize}
\item the Metropolis-Hastings algorithm \citep{hastings1970}. It
  consists in a single-particle sampling of the parameter space. At
  each step, $t$, the proposed parameter vector, $\vec{\theta}'$, is
  extracted from the distribution $q(\vec{\theta}'|\vec{\theta}(t))$,
  centered on $\vec{\theta}(t)$;
\item the stretch-move algorithm \citep{goodman2010}. It represents a
  multi-particle approach.  At each step, $t$, the proposed position
  $\vec{\theta}_i'$ for the i$-th$ particle is located on the line
  connecting $\vec{\theta}_i(t)$ and $\vec{\theta}_j(t)$, where the
  latter is randomly extracted from the particle ensemble.  This
  allows an exchange of information between particles in the cloud.
\end{itemize}

The current version of the CBL implements Gaussian priors, though
minor modifications are required to include different
parameterisations. Examples of scientific results obtained using the
implemented Bayes methods are provided e.g. in \citet{veropalumbo2014,
  veropalumbo2015}.

The functions used to handle the likelihood sampling and the parameter
posteriors are implemented in the classes {\mbox{\color{ForestGreen}
    cosmobl::Prior}}, {\mbox{\color{ForestGreen} cosmobl::Chain}} and
{\mbox{\color{ForestGreen} cosmobl::MCMC}}.


\section{Other functions}
\label{sec:func}

In addition to the classes described above, a large set of generic
functions are included in the {\mbox{\color{ForestGreen} cosmobl}}
namespace. Among them, the set includes: i) functions of generic use,
such as to handle errors and warning messages or endian conversions;
ii) functions to manipulate vectors and matrices; iii) functions for
statistical analyses; iv) functions to calculate distances; v) special
functions (e.g. Legendre polynomials).  A full documentation can be
found at the CBL webpage.


\section{Conclusions}
\label{sec:conc}

We presented the {\small CosmoBolognaLib}, a large set of publicly
available C++ libraries for cosmological calculations. This represents
a {\em living project} with the primary goal of defining a common
numerical environment for cosmological investigations of the
large-scale structure of the Universe. These libraries provide several
classes and methods that can be used to handle catalogues of
extragalactic sources, measure statistical quantities, such as 2PCF
and 3PCF, and derive cosmological constraints. The {\small doxygen}
documentation is provided at the same webpage where the libraries can
be downloaded, together with a set of sample codes that show how to
use this software in either C++ or Python codes.


\section*{Acknowledgments}
We thank C. Fedeli, C. Giocoli, M. Roncarelli and
F. Villaescusa-Navarro for helping in the implementation and
validation of the {\small CosmoBolognaLib}. We are also grateful to
A. Cimatti and L. Moscardini for stimulating discussions about
scientific issues related to these libraries.


\appendix

\section{How to compile}
\label{sec:compile}

To compile the CBL, the following steps have to
be followed:

\begin{enumerate}
\item if not already present in the system, install the following
  external libraries: GSL (GNU Scientific Library)\footnote{
    \url{http://www.gnu.org/software/gsl}}, FFTW\footnote{
    \url{http://www.fftw.org}}, OpenMP\footnote{
    \url{http://openmp.org/}} and Numerical Recipes (Third
  Edition)\footnote{ \url{http://www.nr.com}};
\item download the CosmoBolognaLib.tar archive \footnote{
  \url{http://apps.difa.unibo.it/files/people/federico.marulli3/}} and
  unpack all the files in a folder called CosmoBolognaLib/;
\item enter the CosmoBolognaLib/ folder and type: {\color{blue} \ttfamily make}
  
\end{enumerate}

In this way, the full set of libraries will be compiled, using the GNU
project g++ compiler.

Other Makefile options are the following:
\begin{itemize}
\item {\color{blue} \ttfamily make lib}* $\rightarrow$ compile the *
  library (e.g.  {\color{blue} \ttfamily make libFUNC} will compile
  the library libFUNC.so)
\item {\color{blue} \ttfamily make python} $\rightarrow$ compile the
  Python wrappers
\item {\color{blue} \ttfamily make clean} $\rightarrow$ remove all the
  object files that have been already compiled
\item {\color{blue} \ttfamily make purge} $\rightarrow$ make clean +
  remove all the library files (i.e. *.so)
\item {\color{blue} \ttfamily make purgeALL} $\rightarrow$ make purge
  + remove all the files stored for internal calculations
\item {\color{blue} \ttfamily make CAMB} $\rightarrow$ compile the
  external software CAMB
\item {\color{blue} \ttfamily make CLASS} $\rightarrow$ compile the
  external software CLASS
\item {\color{blue} \ttfamily make CLASSpy} $\rightarrow$ compile the
  Python wrapper for the external software CLASS
\item {\color{blue} \ttfamily make MPTbreeze} $\rightarrow$ compile
  the external software MPTbreeze
\end{itemize}

To compile on MAC OS, the command to be used is: {\color{blue}
  \ttfamily make SYS=`MAC'}. If needed, both the compiler and the
compilation flags can be specified by modifying the variables {\em C}
and {\em FLAGS} defined in the Makefile, e.g. typing: {\color{blue}
  \ttfamily make C=icpc FLAGS=`-O0 -g'} .


\section{Examples}
\label{sec:examples}

In this section, we provide some key examples that can help the user
to understand the main functionalities of the CBL. Many other example
codes are provided at the CBL webpage.


\subsection{Cosmology}
\label{subsec:ex_cosmology}

The following code shows how to use the class
{\mbox{\color{ForestGreen} cosmobl::Cosmology}} to estimate the value
of $f\sigma_8$ at $z=1$. When the object is created, at line 13,
default cosmological parameters are used. Alternatively, the
cosmological parameters can be set directly by using a different
constructor, as shown in the example code provided at the CBL webpage.

\begin{lstlisting}[style=C++, mathescape, caption={Example of how to use the class cosmobl::Cosmology}.]
  // include the header file of libCOSM.so 
  #include "Cosmology.h" 
  
  // the CosmoBolognaLib directory
  const string cosmobl::par::DirCosmo = DIRCOSMO;

  // the current directory
  const string cosmobl::par::DirLoc = DIRL;

  int main () {
    
    // create an object of class Cosmology
    cosmobl::Cosmology cosm;
    
    // set the redshift
    double z = 1.;
    
    // set the external code used to estimate $\sigma_8$
    string method = "CAMB";
    
    // print the value of $f(z)\cdot\sigma_8(z)$
    cout << cosm.fsigma8(z, method) << endl; 
    
    return 0;
  } 
\end{lstlisting}


\subsection{The 2PCF}
\label{subsec:ex_2pt}

The following code illustrates how to handle a catalogue of galaxies
and measure the 2PCF. Firstly, a vector of objects of class
{\mbox{\color{ForestGreen} cosmobl::Galaxy}} is created (lines
21-27). Then, two objects of class {\mbox{\color{ForestGreen}
    cosmobl::Catalogue}} are created, one containing the galaxies
(line 30), and the other containing objects randomly distributed in
the same box (lines 32-35). Finally, the 2PCF is computed by
exploiting the methods of the class {\mbox{\color{ForestGreen}
    cosmobl::TwoPointCorrelation}} (lines 52-64).

\begin{lstlisting}[style=C++, mathescape, caption={Example of how to measure the two-point correlation function}.]
  // include the header file of libRANDOM.so #include
  "RandomCatalogue.h"

  // include the header file of libTWOP.so
  #include "TwoPointCorrelation.h"

  // the CosmoBolognaLib directory
  const string cosmobl::par::DirCosmo = DIRCOSMO;

  // the current directory
  const string cosmobl::par::DirLoc = DIRL;

  int main () {
    
    // create an object of class Cosmology 
    cosmobl::Cosmology cosm;

    // read the coordinates
    string cFile = cosmobl::par::DirLoc+"cat.dat";
    ifstream fin(cFile.c_str());
    vector<shared_ptr<cosmobl::Object>> obj;
    double RA, DEC, RED;
    while (fin >> RA >> DEC >> RED) {
      shared_ptr<cosmobl::Galaxy> gal(new cosmobl::Galaxy{RA, DEC, RED, cosm});
      obj.push_back(gal);
    }
    fin.close();

    // create the object of class Catalogue
    shared_ptr<cosmobl::Catalogue> cat(new cosmobl::Catalogue{obj});

    // create the object of class RandomCatalogue
    int nn = cat->nObjects()*50;
    string rdir = cosmobl::par::DirLoc+"random/";
    auto ran = cosmobl::random_catalogue_box(cat, nn, rdir);

    // minimum separation
    double rMIN = 1.;
    
    // maximum separation
    double rMAX = 50.;
    
    // logarithmic binning size
    double lgbin = 0.05; 

    // linear binning size
    double bin = 0.5; 
    
    // angular binning size
    double cos = 0.02; 
    
    // create the object of class TwoPointCorrelation
    cosmobl::TwoPointCorrelation TwoP {cat, ran};
    
    // set the cosmobl::parameters
    TwoP.setParameters(rMIN, rMAX, lgbin, bin, cos);
    
    // measure the two-point correlation function
    string pdir = cosmobl::par::DirLoc+"pairs/";
    TwoP.measure_xi(pdir);
    
    // store the outputs
    string odir = cosmobl::par::DirLoc+"output/";
    TwoP.write_xi(odir);
    
    return 0;
  }
\end{lstlisting}


\subsection{Python}
\label{subsec:ex_python}

This last example shows how to import the CBL as a module in a Python
code. Specifically, the following code computes the comoving distance
at $z=1$. Only a subset of the CBL has currently a Python wrapper. We
plan to extend this to all the CBL functions in a future version of
the libraries. For now, this example is just meant to illustrate the
potentiality of this approach.

\begin{lstlisting}[style=Python, mathescape, caption={Example of how to use the {\small CosmoBolognaLib} in Python}.]
  # import the module Cosmology of CosmoBolognaLib 
  from CosmoBolognaLib import Cosmology 

  # set the cosmology, using default parameters
  cosm = Cosmology()

  # set the redshift
  z = 1

  # compute the comoving distance
  dc = cosmo.D_C(z)
\end{lstlisting}
  

\bibliographystyle{elsarticle-harv} \bibliography{bib}

\end{document}